\documentclass{article}
\usepackage{spconf,amsmath,graphicx}
\usepackage{xcolor}
\usepackage{balance}

\title{UniCT DMI Solution for 3rd COV19D Competition
on COVID-19 Detection through attention-based CNN for CT Scan}
\makeatletter
\def\@name{\emph{Rondinella A.$^{\star}$, Guarnera F.$^{\star}$, Giudice O.$^{\star}$, Ortis A.$^{\star}$, Rundo F.$^{\dagger}$, Battiato S.$^{\star}$}}
\makeatother

\address{$^{\star}$Department of Mathematics and Computer Science, University of Catania, Catania, Italy\\
$^{\dagger}$STMicroelectronics, ADG R\&D Power and Discretes Division, Catania, Italy  \\\\
\emph{alessia.rondinella@phd.unict.it, francesco.guarnera@unict.it, giudice@dmi.unict.it, }\\
\emph{ortis@dmi.unict.it, francesco.rundo@st.com, battiato@dmi.unict.it}
}
\begin{document}
%\ninept
%
\maketitle
\begin{abstract}
This paper presents our solution for the first challenge of the 3rd Covid-19 competition, which is part of the "AI-enabled Medical Image Analysis Workshop" organized by IEEE International Conference on Acoustic, Speech and Signal Processing (ICASSP) 2023. Our proposed solution is based on a Resnet as a backbone network with the addition of attention mechanisms. The Resnet provides an effective feature extractor for the classification task, while the attention mechanisms improve the model's ability to focus on important regions of interest within the images. We conducted extensive experiments on the provided dataset and achieved promising results. Our proposed approach has the potential to assist in the accurate diagnosis of Covid-19 from chest computed tomography images, which can aid in the early detection and management of the disease.
\end{abstract}
\begin{keywords}
Covid-19 detection, Deep Learning, Computed Tomography scan classification, Medical imaging
\end{keywords}
\section{Introduction}
\label{sec:intro}

The outbreak of the Covid-19 pandemic has had a devastating impact on the world, causing widespread illness and death \cite{ciotti2020covid}. Early detection of the disease is crucial for the timely management and control of the spread of the virus. Medical imaging has emerged as a valuable tool in the fight against Covid-19, particularly in the detection of lung abnormalities associated with the disease \cite{kollias2022ai}. Chest computed tomography (CT) has been widely used in the diagnosis and monitoring of Covid-19 patients, as it provides detailed information about the extent and severity of lung involvement. However, the interpretation of chest CT images is complex, and requires expert knowledge and experience. 
In recent years, deep learning has shown great potential in the field of medical imaging, particularly in the automated detection and diagnosis of various diseases. There has been a rise in research into employing deep learning algorithms to identify and diagnose diseases using medical imaging data, such as chest CT scans, and Covid-19 is no exception.
In this paper, we present our fully automated solution for the first challenge of the 3rd Covid-19 competition \cite{kollias2023ai}, which is part of the "AI-enabled Medical Image Analysis Workshop" organized by IEEE International Conference on Acoustic, Speech and Signal Processing (ICASSP) 2023. There are two challenges in the competition: the COVID-19 Detection Challenge and the COVID-19 Severity Detection Challenge. The first challenge aims to differentiate COVID-19 cases from non-COVID cases by manually annotating each CT scan with respect to COVID-19 and non-COVID-19 categories. The second challenge focuses on determining the severity of COVID-19 by dividing it into four stages, including Mild, Moderate, Severe, and Critical. We propose a framework for the detection of Covid-19 from chest CT scans. Our proposed approach is based on a modified Resnet backbone network with attention levels to focus on important regions of interest within the lung.
However, there are still challenges to be addressed in the use of deep learning for Covid-19 detection, including the need for large and diverse datasets, the potential for bias in the training data, and the interpretability and transparency of the models. Further research is needed to address these challenges and to develop more robust and reliable deep learning models for Covid-19 detection and diagnosis.
The remainder of this paper is organized as follows: Section \ref{sec:sota} provide a brief overview of the existing literature on the use of medical imaging for Covid-19 detection. In Section \ref{sec:methodology} we describe the dataset and present our proposed automated approach for the detection of Covid-19 from chest CT scans. Section \ref{sec:results} present the experimental setup and analysis of our approach, followed by a discussion. Finally, conclusions are summarized in Section \ref{sec:future_work}

\begin{figure*}[t]
\centering
\begin{minipage}[b]{\linewidth}
  \centering
  \centerline{\includegraphics[width=\linewidth]{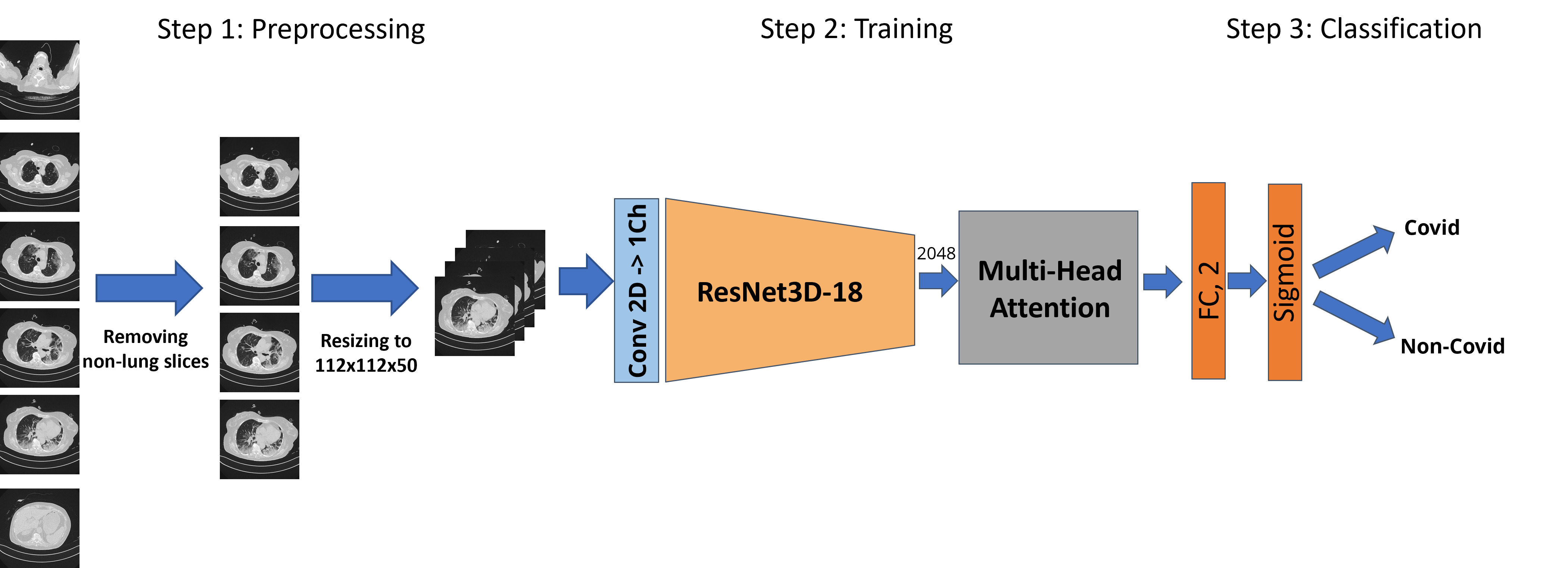}}
%  \vspace{1.5cm}
\end{minipage}
\caption{The proposed pipeline for Covid-19 detection from chest CT scans. First, the non-lung slices were removed, followed by slice resizing. The proposed network was then applied to the preprocessed slices to perform Covid-19 detection.}
\label{fig:pipeline}
\end{figure*}

\section{State of the art}
\label{sec:sota}
Deep learning has emerged as a promising tool for medical image analysis, with applications in both Covid-19 detection and general medical image analysis. In general medical image analysis, deep learning has been used for various tasks, such as periprosthetic joint infection (PJI) classification \cite{guarnera2023early} and multiple sclerosis (MS) image segmentation \cite{rondinella2023boosting}.
The Covid-19 pandemic has led to a large demand for effective and efficient methods for the detection and diagnosis of the disease. Deep learning approaches have emerged as a promising tool for this task, owing to their ability to learn complex representations of medical images and extract relevant features for accurate detection \cite{kollias2020deep}\cite{kollias2020transparent}\cite{kollias2018deep}.
Several studies have been conducted to explore the potential of deep learning in Covid-19 detection. For instance, authors in \cite{napoli2022mixup} propose a mixup data augmentation tecnique on Covid-19 to dedicing the severity of infection at slice-level. In \cite{jain2020deep} Jain et al. proposed a deep learning framework to distinguish COVID-19 induced pneumonia on chest X-ray pictures from healthy patients. Similarly, \cite{bhattacharyya2022deep} propose a framework that first segments the lungs from X-ray images and then classifies them into COVID-19, pneumonia and normal lung.
Authors in \cite{mukherjee2021deep} use a Convolutional Neural Network (CNN) to detect the presence of Covid-19 on both X-ray and CT scan images.
In \cite{wang2020fully} the authors propose a deep learning approach to automatically focused on abnormal areas of lungs to detect Covid-19. Similarly, in \cite{polsinelli2020light} Polsinelli et al. propose a CNN that follow the SqueezeNet architecture to classify chest CT scan images into Covid-19, healty or pneumonia.

\section{Methodology}
\label{sec:methodology}

Chest CT scans play a crucial role in the detection and diagnosis of Covid-19, as they can provide detailed information about the extent and severity of lung involvement in patients. In the proposed framework for the automated detection of Covid-19 from chest CT scans, the specific preprocessing steps and the modified Resnet backbone network with attention modules will be described in detail in Section \ref{sec:implementation}

\subsection{Dataset}
\label{ssec:dataset}

The COV19-CT-DB dataset \cite{arsenos2022large}\cite{kollias2021mia} was used to evaluate the performance of our proposed approach for automated Covid-19 detection from chest CT scans. This dataset consists of 3D CT scans of the chest labelled by experts, corresponding to a large number of patients. In total, the dataset contains 3746 CT scans, with 1147 scans labelled as Covid and 2599 scans labelled as non-Covid. The scans were then divided into training and validation sets, with 3032 scans used for training and 714 scans used for validation.
This dataset was used for both challenges of the 3rd Covid-19 competition. For the first challenge, the dataset was divided into Covid and non-Covid labels. For the second challenge, the same dataset was used, but with the Covid scans labelled according to four levels of severity: medium, moderate, severe, and critical.

\subsubsection{Pre-processing}
\label{sssec:preprocessing}

The COV19-CT-DB dataset was preprocessed to extract only the relevant scans for the Covid-19 classification task. Since not all scans in the dataset have the same length, we decided to select only those scans containing the lung region, which is the primary region of interest for Covid-19 detection.

To extract these scans, we selected 50 consecutive scans per CT scan, starting from the center of the CT. This allowed us to capture a consistent and representative portion of the lung region in each scan. We also ensured that each selected scan had the same resolution of $112 \times 112$ pixels.

\subsection{Implementation details}
\label{sec:implementation}
In this paper, we propose a fully automated approach for the detection of Covid-19 from chest CT scans. Our approach is based on a modified 3D Resnet as a backbone, which is a widely used deep learning architecture for image classification tasks. Specifically, we used the 3D Resnet-18 architecture as our backbone \cite{tran2018closer}, which consists of 18 layers and is well-suited for processing 3D medical images.

To improve the performance of the baseline architecture for Covid-19 detection, we added attention mechanisms to the network. In particular, we added four heads of multi-head attention, which allowed the network to focus on important regions of interest within the lung. The attention mechanism enhances the network's ability to identify and discriminate between Covid-19 and non-Covid-19 cases, by focusing on the regions of the lung that are most indicative of the disease.

\section{Experiments and results}
\label{sec:results}

\subsection{Experimental setup}
\label{sec:setup}
As explained in Section \ref{sec:methodology} we trained our proposed network on the preprocessed COV19-CT-DB dataset, using a binary cross-entropy with logits loss function and the Adam optimizer with a learning rate of 0.0001. We used a batch size of 4 and trained the network for 50 epochs.
To evaluate the performance of our proposed approach, we used the validation set from the challenge dataset.
We report the results in terms of Recall, Precision and Macro F1 score, which are commonly used evaluation metrics for binary classification tasks.
Figure \ref{fig:pipeline} shows the overall proposed pipeline.
%We report the results in terms of Macro F1 score, which is commonly used evaluation metrics for binary classification tasks.

\subsection{General results}
\label{sec:quantitative_results}

We evaluated the performance of our proposed approach on the validation set of the COVID-19 detection challenge using various evaluation metrics. Table \ref{table:ablation} shows the results of the baseline approach compared with the 3D Resnet-18 model and our proposed approach with attention mechanisms.
The baseline architecture utilizes a ResNet combined with a Recurrent Neural Network (RNN) and achieves a macro F1 score of 0.77.
The current approach builds upon previous research efforts in developing deep neural architectures for predicting COVID19, following the work of \cite{kollias2020deep}\cite{kollias2020transparent}. 
As can be seen from the table, our proposed approach with attention mechanisms outperformed the baseline model in terms of F1 score. Specifically, the network with attention achieved a Recall of 0.88, a Precision of 0.92 and a macro F1 score of 0.90, indicating its high performance in detecting Covid-19 from chest CT scans.
Overall, our proposed approach provides a promising solution for the automated detection of Covid-19 from chest CT scans, which can be useful in the early detection and management of the disease.
\begin{table}[t]
\centering
\resizebox{\linewidth}{!}{
\begin{tabular}{c||c||c||c}
\hline
\hline
\textbf{Architecture}                                    & \textbf{Recall} & \textbf{Precision} & \textbf{Macro F1 Score} \\ \hline
Baseline\cite{kollias2023ai} & -      & -      & 0.74   \\
ResNet3D-18\cite{tran2018closer} & 0.8847      & 0.9138      & 0.8973   \\
ResNet3D-18 + MHA                           & \textbf{0.8867}      & \textbf{0.9233} & \textbf{0.9021}   \\ \hline
\hline
\end{tabular}
}
\caption{Ablation studies with different network configurations employing the validation set of COV19-CT-DB.}
\label{table:ablation}
\end{table}

\section{Conclusion and future works}
\label{sec:future_work}

In this paper, we presented a fully automated approach for the detection of Covid-19 from chest CT scans using deep learning techniques. Our proposed framework utilizes a modified Resnet backbone network with attention mechanisms to detect Covid-19 from 3D CT scans of the chest. The COV19-CT-DB dataset, consisting of 3D CT scans of the chest labeled as Covid and non-Covid, was used to train and validate our proposed approach. The experimental results show that our approach achieves a macro F1 score of 0.90, which is promising for Covid-19 detection.
Future research will focus on investigating the potential of our framework on a larger dataset, as well as exploring the possibility of incorporating other imaging modalities for Covid-19 detection. Furthermore, we plan to investigate the potential of transfer learning to improve the performance of our proposed approach. Finally, we will also explore the use of our framework for other respiratory diseases that affect the lungs..

% Below is an example of how to insert images. Delete the ``\vspace'' line,
% uncomment the preceding line ``\centerline...'' and replace ``imageX.ps''
% with a suitable PostScript file name.
% -------------------------------------------------------------------------

\section*{Acknowledgements}
\label{sec:acknowledgements}
Alessia Rondinella is a PhD candidate enrolled in the National PhD in Artificial Intelligence, XXXVII cycle, course on Health and life sciences, organized by Università Campus Bio-Medico di Roma.

Experiments were carried out thanks to the hardware and software granted and managed by iCTLab S.r.l. - Spinoff of University of Catania.

%\vfill\pagebreak

\bibliographystyle{IEEEbib}
\bibliography{refs}

\end{document}